\documentclass[aps,prd,nofootinbib,10pt]{revtex4-2}

\usepackage{graphicx}
\usepackage{bm}
\usepackage{mathrsfs}
\usepackage{amsthm,amsmath,amssymb}
\usepackage{verbatim}
\usepackage{siunitx}
\usepackage{booktabs}
\usepackage[colorlinks,linkcolor=red,anchorcolor=green,citecolor=blue]{hyperref}

\begin{document}

\title{Stationary States for Fermions in an External Electric Field}

\author{Xuan Zhao}
\email{zhaox21@mails.tsinghua.edu.cn}
\affiliation{Department of Physics, Tsinghua University, Beijing 100084, China}

\author{Yi Wang}
\email{y-wang23@mails.tsinghua.edu.cn}
\affiliation{Department of Physics, Tsinghua University, Beijing 100084, China}

\author{Pengfei Zhuang}
\email{zhuangpf@mail.tsinghua.edu.cn}
\affiliation{Department of Physics, Yantai University, Yantai 264005, China}
\affiliation{Department of Physics, Tsinghua University, Beijing 100084, China}
\affiliation{Southern Center for Nuclear-Science Theory (SCNT), Institute of Modern Physics, Chinese Academy of Sciences, Huizhou 516000, China}

%\date{\today}

\begin{abstract}
We present a relativistic analysis of fermions in an external electric field by non-perturbatively solving the Dirac equation with a static gauge. Different from the magnetic field effect, the fermion wave function in an electric field oscillates asymptotically, which results in the absence of bound states in an infinite system. For a confined fermion, the confinement is gradually canceled by the electric field, and the fermion becomes deconfined when the electric coupling is stronger than the confinement coupling. However, a fermion in an electric field can be confined to a finite system by applying the MIT bag boundary condition, namely, the disappearing normal component of the probability current at the boundary. The solutions obtained can serve as a basis for calculating dynamical processes in the presence of a strong electric field, such as those occurring in the early stage of relativistic heavy-ion collisions, where an extremely strong electric field is expected to be generated.
\end{abstract}

\maketitle

%%%%%%%%%%%%%%%%%%%%%%%%%%%%%%%%%%%%%%%%%%%%%%
\section{Introduction}
%%%%%%%%%%%%%%%%%%%%%%%%%%%%%%%%%%%%%%%%%%%%%%
Fermionic systems in external electromagnetic fields have long been an important topic in quantum electrodynamics (QED) and quantum chromodynamics (QCD). In recent heavy ion collisions at high energies, the strongest electromagnetic fields in nature are expected to be created by the spectators of the collisions and the induced current in the quark gluon plasma, with field strength reaching or even exceeding the characteristic QCD scale~\cite{Tuchin:2014hza,Voronyuk:2011jd,Deng:2012pc,Bzdak:2011yy}. While both electric and magnetic fields play important roles in the collisions, most research works focused on the magnetic field induced phenomena, such as chiral magnetic effect~\cite{Kharzeev:2007jp,Fukushima:2008xe,Kharzeev:2015znc,Huang:2018wdl}, magnetic catalysis and inverse magnetic catalysis~\cite{Fukushima:2012kc,Mei:2024rjg,Yu:2014xoa,Fraga:2013ova,Cao:2021rwx}, and polarization difference between hyperons and anti-hyperons~\cite{Guo:2019joy}. A widely studied phenomenon related to electric field effects is the Schwinger pair production~\cite{Schwinger:1951nm,Kluger:1992gb,Hebenstreit:2010vz,Soldati:2011gi,Sheng:2018jwf,Sheng:2019ujr,Cao:2015dya,Yamamoto:2012bd}, which can well be described through the solutions of the time-dependent Dirac equation~\cite{Nikishov:1969tt,Soldati:2011gi}. To study the electric field effect on the static properties of fermions, such as mass and binding energies~\cite{Cao:2015cka,Wang:2018gmj,Cao:2019hku,Cao:2020pjq,Cao:2015dya,Su:1984gq,Payod:2024fsm,Xiang:2024wxh,Chen:2020xsr,Biaogang:2019hij}, one needs to know the stationary solutions of the Dirac equation in an external electric field.

It is widely accepted that a charged fermion cannot be in a bound state in an infinite uniform electric field. This can be understood with simple classical arguments based on electric acceleration, but an explicit and rigorous proof should come from the stationary solution of the Dirac equation. The situation becomes more nontrivial when a confinement interaction is included, which can be realized in relativistic heavy ion collisions where quarks are confined in hadrons when the temperature of the system is low. How does the electric field compete with such confinement mechanism and under what kinds of conditions the bound state disappears are questions that can only be answered by the Dirac equation with electric and confinement potentials.

In ultra-peripheral heavy ion collisions, there is no strong (QCD) interaction, the electromagnetic processes such as light-by-light scattering and Breit–Wheeler pair production are successfully described within the perturbative QED using an equivalent photon picture~\cite{ATLAS:2017fur, ATLAS:2019azn, STAR:2019wlg}. In contrast, for central and semi-central heavy ion collisions, the electromagnetic field acts as a classical background field influencing the QCD and QED dynamics of relativistic fermions produced at early times~\cite{Chen:2024lmp, Chen:2025hpm}. In this case, a description based on non-perturbative solutions of the Dirac equation in an external field becomes necessary. Although the space–time profile of the electromagnetic field in heavy ion collisions is highly inhomogeneous, it can be approximated as a local constant within a sufficiently small space–time cell, allowing one to integrate the corresponding local cross section over the entire space–time evolution of the collision, as implemented for example in the Anomalous-Viscous Fluid Dynamics model~\cite{Shi:2019wzi}.

In this work, we address stationary solutions of the Dirac equation under an external electric field, especially the bound states, by providing a non-perturbative analysis in three cases: an infinite system of fermions interacting with only the electric field, an infinite system including a confinement interaction, and a finite system with a suitable boundary condition. The paper is organized as follows. In Section~\ref{sec2}, we solve the Dirac equation with a static gauge for a uniform electric field. We will prove explicitly the disappearance of a bound state in this case. Section~\ref{sec3} is devoted to the competition between the confinement and electric interactions, leading to a transition from confinement to deconfinement when the electric field is strong enough. In Section~\ref{sec4}, we derive the energy quantization for a confined fermion in a finite system via taking the MIT bag boundary condition. A summary and possible applications are given in Section~\ref{sec5}.

Throughout this paper, we adopt the natural system of units  with $\hbar=c=1$ and the metric convention $g_{\mu \nu}= \mathrm{diag}(+,-,-,-)$. The coordinate four-vectors are denoted as $x^{\mu}=(t,\mathbf x)=(t,x,y,z)$.

%%%%%%%%%%%%%%%%%%%%%%%%%%%%%%%%%%%%%%%%%%%%%%
\section{Stationary solution in an external electric field}
\label{sec2}
%%%%%%%%%%%%%%%%%%%%%%%%%%%%%%%%%%%%%%%%%%%%%%
For fermions moving in a constant electric field directed along the $z$-axis, the Dirac equation reads
\begin{equation}
\left(i\gamma^\mu D_\mu-m\right)\psi(t,{\mathbf x})=0
\label{Dirac}
\end{equation}
with fermion mass $m$, electric charge $q$ and covariant derivative $D_\mu=\partial_\mu+iqA_\mu$. For a time-dependent potential $A^\mu=(0,0,0,Et)$ with $E$ being the strength of the external field, the solution of the equation and its application to pair production are systematically studied in literature such as \cite{Nikishov:1969tt,Kluger:1992gb,Hebenstreit:2010vz,Soldati:2011gi,Sheng:2018jwf,Sheng:2019ujr}. Here we focus on the stationary solution of the equation with a static potential $A^{\mu}=(Ez,0,0,0)$. The $\gamma$ matrices are defined in the standard Dirac representation.

Since the Hamiltonian
\begin{equation}
\label{hamil}
\hat H = -i\gamma^0{\bm \gamma}\cdot{\bm \nabla}+|qE|z+m\gamma^0
\end{equation}
is time independent, and the momenta $p_x$ and $p_y$ in the $x$- and $y$-directions are conserved, the stationary solution can be factorized as
\begin{equation}
\psi(t,{\mathbf x}) = e^{-i(p_0 t-p_x x-p_y y)}\phi(z),
\end{equation}
where $p_0$ is the particle energy, and the spinor $\phi(z)$ is a function of the variable $z$ and satisfies the equation
\begin{equation}
\left[(p_0-|qE|z)\gamma^0-p_x\gamma^1-p_y\gamma^2+i\gamma^3\partial_z-m\right]\phi(z)=0.
\end{equation}
We further separate the coordinate space and spin space of the wave function,
\begin{equation}
\phi(z) = \left[(p_0-|qE|z)\gamma^0-p_x\gamma^1-p_y\gamma^2+i\gamma^3\partial_z+m\right]f(z)\chi,
\end{equation}
where $f(z)$ is a complex scalar function of $z$, it and the bispinor $\chi$ are controlled by the equation
\begin{equation}
\label{fchi}
\left[\frac{d^2}{dz^2}+(p_0-|qE|z)^2-p_\bot^2-m^2+i|qE|\gamma^0\gamma^3\right]f(z)\chi=0,
\end{equation}
where $p_\bot^2=p_x^2+p_y^2$ is the transverse momentum square. This is still a coupled equation for $f$ and $\chi$. To make the separation successful, we take explicitly the four independent bispinors as $\chi_\epsilon^s\ (\epsilon=\pm, s=\pm)$ in the Dirac representation,
\begin{equation}
\chi_+^+=\frac{1}{\sqrt{2}}
\begin{pmatrix}
1 \\
0 \\
1 \\
0 \\
\end{pmatrix},\ \ \ \ \
\chi_+^-=\frac{1}{\sqrt{2}}
\begin{pmatrix}
0 \\
1 \\
0 \\
-1 \\
\end{pmatrix},\ \ \ \ \
\chi_-^+=\frac{1}{\sqrt{2}}
\begin{pmatrix}
1 \\
0 \\
-1 \\
0 \\
\end{pmatrix},\ \ \ \ \
\chi_-^-=\frac{1}{\sqrt{2}}
\begin{pmatrix}
0 \\
1 \\
0 \\
1 \\
\end{pmatrix}.
\label{chi}
\end{equation}
The bispinors chosen in this way are common eigenstates of the spin component along the $z$-axis $\Sigma_z=i\gamma^1\gamma^2/2$ and the operator $\gamma^0\gamma^3$ in the equation (\ref{fchi}),
\begin{equation}
\Sigma_z\chi_\epsilon^s=\frac{s}{2}\chi_\epsilon^s,\ \ \ \ \ \gamma^0\gamma^3\chi_\epsilon^s=\epsilon\chi_\epsilon^s.
\end{equation}
The four bispinors satisfy the orthonormalization and completeness conditions,
\begin{equation}
\sum_{\alpha=1}^4(\chi_\epsilon^{s\dagger})^\alpha(\chi^{s'}_{\epsilon'})_\alpha=\delta_{ss'}\delta_{\epsilon\epsilon'},\ \ \ \ \
\sum_{\epsilon,s}(\chi_\epsilon^{s\dagger})^\alpha(\chi^s_\epsilon)_\beta=\delta^\alpha_\beta.
\end{equation}
Therefore, the four independent wave functions $\phi_\epsilon^s(z)$ can be expressed as
\begin{equation}
\phi_\epsilon^s(z) = \left[(p_0-|qE|z)\gamma^0-p_x\gamma^1-p_y\gamma^2+i\gamma^3\partial_z+m\right]f_\epsilon(z)\chi_\epsilon^s
\end{equation}
with $f_\epsilon$ characterized by the normal differential equation
\begin{equation}
\left[\frac{d^2}{dz^2}+(p_0-|qE|z)^2-p_\bot^2-m^2 + i\epsilon|qE|\right]f_\epsilon(z)=0.
\end{equation}

By introducing the dimensionless variable
\begin{equation}
    \xi(z)=\sqrt{|qE|}\left(z-\frac{p_0}{|qE|}\right)
\end{equation}
and the dimensionless parameter
\begin{equation}
    \lambda=\frac{p_\bot^2+m^2}{|qE|},
\end{equation}
the differential equation for $f_\epsilon$ is simplified as
\begin{equation}
\label{fxi}
    \left(\frac{d^2}{d\xi^2}+\xi^2-\lambda+i\epsilon\right)f_\epsilon(\xi)=0.
\end{equation}
The solution of the equation is the parabolic cylinder function $D_\nu(z)$~\cite{gradshteyn2014table}, and there are two linearly independent solutions for a fixed $\epsilon$,
\begin{eqnarray}
&& f_+(\xi|\pm)=D_{i\lambda/2}(\pm\xi_+),\nonumber\\
&& f_-(\xi|\pm)=D_{i\lambda/2-1}(\pm\xi_+),\ \ \ \ \ \xi_\pm(z) = (1\pm i)\xi(z).
\end{eqnarray}
The linear independence of the two solutions $D_{i\lambda/2}(\pm\xi_+)$ or $D_{i\lambda/2-1}(\pm\xi_+)$ can be verified by computing the Wronskian~\cite{bateman1953higher}, as detailed in the Appendix.

Taking into account the relations for the bispinors,
\begin{equation}
    \gamma^0\chi_\epsilon^s=\chi_{-\epsilon}^s,\ \ \ \ \ \gamma^1\chi_\epsilon^s=\epsilon s\chi_\epsilon^{-s},\ \ \ \ \ \gamma^2\chi_\epsilon^s=i\epsilon\chi_\epsilon^{-s},\ \ \ \ \ \gamma^3\chi_\epsilon^s=\epsilon\chi_{-\epsilon}^s
\end{equation}
and the recursion formulas for the parabolic cylinder functions~\cite{gradshteyn2014table}, see the Appendix, we eventually find the wave functions $\phi_\epsilon^s(z|\pm)$,
\begin{eqnarray}
\phi_+^+(z|\pm) &=& \left[(p_0-|qE|z)\gamma^0-p_x\gamma^1-p_y\gamma^2+i\partial_z\gamma^3+m\right]D_{i\lambda/2}(\pm\xi_+)\chi_+^+\nonumber\\
&=& \pm(i-1)(i\lambda/2)\sqrt{|qE|}D_{i\lambda/2-1}(\pm\xi_+)\chi_-^+ - p_+D_{i\lambda/2}(\pm\xi_+)\chi_+^- + mD_{i\lambda/2}(\pm\xi_+)\chi_+^+,\nonumber\\
\phi_+^-(z|\pm) &=& \left[(p_0-|qE|z)\gamma^0-p_x\gamma^1-p_y\gamma^2+i\partial_z\gamma^3+m\right]D_{i\lambda/2}(\pm\xi_+)\chi_+^-\nonumber\\
&=& \pm(i-1)(i\lambda/2)\sqrt{|qE|}D_{i\lambda/2-1}(\pm\xi_+)\chi_-^- + p_- D_{i\lambda/2}(\pm\xi_+)\chi_+^+ + mD_{i\lambda/2}(\pm\xi_+)\chi_+^-,\nonumber\\
\phi_-^+(z|\pm) &=& \left[(p_0-|qE|z)\gamma^0-p_x\gamma^1-p_y\gamma^2+i\partial_z\gamma^3+m\right]D_{i\lambda/2-1}(\pm\xi_+)\chi_-^+\nonumber\\
&=& \pm(i-1)\sqrt{|qE|}D_{i\lambda/2}(\pm\xi_+)\chi_+^+ + p_+ D_{i\lambda/2-1}(\pm\xi_+)\chi_-^- + mD_{i\lambda/2-1}(\pm\xi_+)\chi_-^+,\nonumber\\
\phi_-^-(z|\pm) &=& \left[(p_0-|qE|z)\gamma^0-p_x\gamma^1-p_y\gamma^2+i\partial_z\gamma^3+m\right]D_{i\lambda/2-1}(\pm\xi_+)\chi_-^-\nonumber\\
&=& \pm(i-1)\sqrt{|qE|}D_{i\lambda/2}(\pm\xi_+)\chi_+^- - p_- D_{i\lambda/2-1}(\pm\xi_+)\chi_-^+ + mD_{i\lambda/2-1}(\pm\xi_+)\chi_-^-,
\end{eqnarray}
where $p_\pm$ is defined as $p_\pm=p_x\pm i p_y$.

Note that $\epsilon=\pm$ and $s=\pm$ are quantum numbers used to label the wave functions $\phi_\epsilon^s(z|\pm)$ or
\begin{equation}
\psi_\epsilon^s(t,{\mathbf x}|\pm) = \phi_\epsilon^s(t,{\mathbf x}|\pm)e^{-i(p_0 t-p_xx-p_yy)},
\end{equation}
obviously, the labels $\epsilon$ and $s$ in the wave function $\psi_\epsilon^s$ do not, in general, represent the eigenstates of positive/negative energy and spin component,
\begin{equation}
\hat H \psi_\epsilon^s \neq \epsilon p_0\psi_\epsilon^s,\ \ \ \ \ \Sigma_z\psi_\epsilon^s \neq {\frac{s}{2}}\psi_\epsilon^s.
\end{equation}

Using the explicit expression of the bispinors $\chi_\epsilon^s$, the wave functions $\phi_\epsilon^s$ can be written down in a matrix form, for instance,
\begin{eqnarray}
\phi_-^+(z|\pm) &=&\frac{1}{\sqrt{2}}
        \begin{pmatrix}
        \pm(i-1)\sqrt{|qE|}D_{i\lambda/2}(\pm\xi_+)+mD_{i\lambda/2-1}(\pm\xi_+) \\
        p_+D_{i\lambda/2-1}(\pm\xi_+) \\
        \pm(i-1)\sqrt{|qE|}D_{i\lambda/2}(\pm\xi_+)-mD_{i\lambda/2-1}(\pm\xi_+) \\
        p_+D_{i\lambda/2-1}(\pm\xi_+) \\
       \end{pmatrix},\nonumber\\
\phi_-^-(z|\pm) &=&\frac{1}{\sqrt{2}}
       \begin{pmatrix}
       -p_-D_{i\lambda/2-1}(\pm\xi_+) \\
       \pm (i-1)\sqrt{|qE|}D_{i\lambda/2}(\pm\xi_+)+mD_{i\lambda/2-1}(\pm\xi_+) \\
       p_-D_{i\lambda/2-1}(\pm\xi_+) \\
       \pm(1-i)\sqrt{|qE|}D_{i\lambda/2}(\pm\xi_+)+mD_{i\lambda/2-1}(\pm\xi_+) \\
       \end{pmatrix}.
\end{eqnarray}

We now have eight wave functions $\psi_\epsilon^s(t,{\mathbf x}|\pm)$, but there should be only four of them to be independent. In fact, the set with $\epsilon = -$ can be linearly expressed in terms of the set with $\epsilon = +$,
\begin{eqnarray}
\psi_-^+(t,{\mathbf x}|\pm) &=& \pm{1\over (i-1)(i\lambda/2)\sqrt{|qE|}}\left(m\psi_+^+(t,{\bf x}|\pm)+p_+\psi_+^-(t,{\mathbf x}|\pm)\right),\nonumber \\
\psi_-^-(t,{\mathbf x}|\pm) &=& \pm{1\over (i-1)(i\lambda/2)\sqrt{|qE|}}\left(m\psi_+^-(t,{\bf x}|\pm)-p_-\psi_+^+(t,{\mathbf x}|\pm)\right).
\end{eqnarray}
This means that only one of the two sets is linearly independent and forms a complete representation of the solutions of the Dirac equation (\ref{Dirac}). In the following, we will take $\psi_-^s(t,{\mathbf x}|\pm)$ as the four independent solutions of the equation.

A motivation for studying the stationary solution of the Dirac equation in an external electric field is to see if there can exist a bound state in the field. If yes, there will be fermion energy quantization $p_0=E_n$ controlled by some quantum number $n$, like the case in an external magnetic field~\cite{Melrose:1983svt,Kostenko:2018cgv,Kostenko:2019was}. To this end, we employ the asymptotic behavior for a general parabolic cylinder function $D_\nu(z)$ in the limit of $|z|\gg 1$ and $|z|\gg |\nu|$~\cite{gradshteyn2014table}, summarized in the Appendix, which leads to the simplification of those parabolic cylinder functions that appear in the wave functions,
\begin{eqnarray}
&& D_{i\lambda/2}(+\xi_+) = \Delta_1(\xi)=e^{-\pi\lambda/8}(2\xi^2)^{i\lambda/4}e^{-i\xi^2/2},\nonumber\\
&& D_{i\lambda/2}(-\xi_+) = e^{\pi\lambda/2}\Delta_1(\xi),\nonumber\\
&& D_{i\lambda/2-1}(+\xi_+) = 0,\nonumber\\
&& D_{i\lambda/2-1}(-\xi_+) = \Delta_2(\xi)=\frac{\sqrt{2\pi}}{\Gamma(1-i\lambda/2)}e^{\pi\lambda/8}(2\xi^2)^{-i\lambda/4}e^{i\xi^2/2}
\end{eqnarray}
in the limit of $\xi\to\sqrt{|qE|}z, z\to +\infty$, and in turn the asymptotic behavior of the four independent wave functions $\phi_-^s(z|\pm)$,
\begin{eqnarray}
\phi_-^+(z\to+\infty|+) &=& (i-1)\sqrt{|qE|}\Delta_1(\xi)\chi_+^+,\nonumber\\
\phi_-^+(z\to+\infty|-) &=& -(i-1)\sqrt{|qE|}e^{\pi\lambda/2}\Delta_1(\xi)\chi_+^+  +p_+\Delta_2(\xi)\chi_-^- +m\Delta_2(\xi)\chi_-^+,\nonumber\\
\phi_-^-(z\to+\infty|+) &=& (i-1)\sqrt{|qE|}\Delta_1(\xi)\chi_+^-,\nonumber\\
\phi_-^-(z\to+\infty|-) &=& -(i-1)\sqrt{|qE|}e^{\pi\lambda/2}\Delta_1(\xi)\chi_+^- -p_-\Delta_2(\xi)\chi_-^+ +m\Delta_2(\xi)\chi_-^-.
\end{eqnarray}

Note that, while the wave functions $\phi_-^s(z|+)$ become now the eigenstates of the spin operator,
\begin{equation}
\Sigma_z\psi_-^s(z\to+\infty|+)=\frac{s}{2}\psi_-^s(z\to+\infty|+),
\end{equation}
the other two $\psi_-^s(z|-)$ are still combinations of the spin-up and spin-down states. Only in the case without transverse motion $p_\pm =0$, they become the spin eigenstates too,
\begin{equation}
\Sigma_z\psi_-^s(z\to+\infty|-)=\frac{s}{2}\psi_-^s(z\to+\infty|-),\ \ \ \ \ p_\pm=0.
\end{equation}
The analysis for the asymptotic behavior of $\psi_-^s$ in the limit of $\xi\to\sqrt{|qE|}z, z\rightarrow -\infty$ is similar.

The electric potential in the Dirac equation $V(z)=|qE|z$ approaches $+\infty$ at $z\rightarrow +\infty$, we expect to find a solution that vanishes asymptotically. A general solution of the Dirac equation is a linear combination of the four independent wave functions $\phi_-^s(z|\pm)$,
\begin{eqnarray}
\phi(z) &=& C_1\phi_-^+(z|+)+C_2\phi_-^+(z|-)+C_3\phi_-^-(z|+)+C_4\phi_-^-(z|-)\nonumber\\
&\stackrel{z\to+\infty}\longrightarrow& 0.
\end{eqnarray}
The four equations to determine the constants $C_i\ (i=1,2,3,4)$ in the limit of $z\to\ +\infty$ can be explicitly written down as
\begin{eqnarray}
&& (i-1)\sqrt{|qE|}\left(C_1-e^{\pi\lambda/2}C_2\right)\Delta_1(\xi)+\left(mC_2-p_-C_4\right)\Delta_2(\xi)=0,\nonumber\\
&& (i-1)\sqrt{|qE|}\left(C_3-e^{\pi\lambda/2}C_4\right)\Delta_1(\xi)+\left(mC_4+p_+C_2\right)\Delta_2(\xi)=0,\nonumber\\
&& (i-1)\sqrt{|qE|}\left(C_1-e^{\pi\lambda/2}C_2\right)\Delta_1(\xi)-\left(mC_2-p_-C_4\right)\Delta_2(\xi)=0,\nonumber\\
&& -(i-1)\sqrt{|qE|}\left(C_3-e^{\pi\lambda/2}C_4\right)\Delta_1(\xi)+\left(mC_4+p_+C_2\right)\Delta_2(\xi)=0.
\end{eqnarray}
Since $\Delta_1(\xi)$ and $\Delta_2(\xi)$ are independent of each other and do not vanish even in the limit of $z\to +\infty$, these equations are equivalent to
\begin{eqnarray}
&& C_1-e^{\pi\lambda/2}C_2=0,\nonumber\\
&& mC_2-p_-C_4=0,\nonumber\\
&& C_3-e^{\pi\lambda/2}C_4=0,\nonumber\\
&& p_+C_2+mC_4=0.
\end{eqnarray}
The determinant of the linear equations is not zero,
\begin{equation}
\left|\begin{matrix}
	1&  -e^{\pi\lambda/2}&  0&                  0  \\
	0&                  m&  0&               -p_-  \\
	0&                  0&  1&  -e^{\pi\lambda/2}  \\
	0&                p_+&  0&                  m  \\
\end{matrix}\right| = |qE|\lambda \neq 0,
\end{equation}
which implies that it is impossible to have a non-trivial bound state in a constant electric field. This proves strictly the known phenomenon of no bound state for a charged fermion in an electric field due to the electric acceleration.

%%%%%%%%%%%%%%%%%%%%%%%%%%%%%%%%%%%%%%%%%%%%%%
\section{Electric field effect on confined fermions}
\label{sec3}
%%%%%%%%%%%%%%%%%%%%%%%%%%%%%%%%%%%%%%%%%%%%%%
The calculation of the last section shows the impossibility of deriving a bound state in a uniform electric field. A natural question is then the following: If we apply an electric field to a confined fermion, will it enhance or reduce the confinement? To answer this question, we consider a system with both electric potential and confinement potential,
\begin{equation} \label{Dirac4}
  \left[i\gamma^\mu\partial_\mu-g_vz\gamma^0-(m+g_s|z|)\right]\psi(t,{\mathbf x})=0,
\end{equation}
where we have taken the confinement force in the same direction as the electric field, in order to see directly the competition between the two interactions. For simplicity, we considered a linear potential $V_s=g_s|z|$ with coupling $g_s>0$. Note that the two interactions do not work in the same channel, they have different Lorentz structures: the electric potential
$V_v = g_v z$ with coupling $g_v = |qE| > 0$ corresponds to the time component of a four-vector interaction, whereas the confinement potential $V_s = g_s |z|$ is a scalar.
This difference is crucial, as the interplay between the two potentials needs a careful relativistic treatment. After operating on both sides of equation (\ref{Dirac4}) with $i\gamma^\mu\partial_\mu-g_vz\gamma^0+m+g_s|z|$ and performing the variable separation, $\psi(t,{\mathbf x})=e^{-i(p_0 t-p_xx-p_yy)}\varphi(z)$, we obtain the second-order Dirac equation
\begin{equation}
\label{varphi}
\left[{d^2\over dz^2}+(p_0-g_vz)^2-(m\pm g_sz)^2-p_\bot^2-ig_v\gamma^3\gamma^0\mp ig_s\gamma^3\right]\varphi^\pm(z)=0
\end{equation}
for
\begin{equation}
\varphi(z)=\left\{\begin{array}{ll}
\varphi^+(z)\quad & z>0\\
\varphi^-(z)\quad & z<0
\end{array}\right..
\end{equation}

We first consider a stronger scalar potential $V_s$ in comparison with the vector potential $V_v$ with $g_s > g_v$. To diagonalize the matrix $-ig_v\gamma^3\gamma^0\mp ig_s\gamma^3$ in the equation (\ref{varphi}), we take further a transformation for the components $\varphi^\pm_i(z)\ (i=1,2,3,4)$,
\begin{equation}
\label{trans}
    \begin{pmatrix}
        \varphi^\pm_1 \\
        \varphi^\pm_2 \\
        \varphi^\pm_3 \\
        \varphi^\pm_4 \\
    \end{pmatrix}
    =
    \begin{pmatrix}
        0 & iP_\pm & 0 & -iP_\pm \\
        -iP_\pm & 0 & iP_\pm & 0 \\
        0 & 1 & 0 & 1 \\
        1 & 0 & 1 & 0 \\
    \end{pmatrix}
    \begin{pmatrix}
        \tilde\varphi^\pm_1 \\
        \tilde\varphi^\pm_2 \\
        \tilde\varphi^\pm_3 \\
        \tilde\varphi^\pm_4 \\
    \end{pmatrix}
\end{equation}
with $P_\pm = \sqrt{g_s^2- g_v^2}/(g_v\pm g_s )$, which leads to the normal differential equations,
\begin{eqnarray}
&& \left[{d^2\over dz^2}+(p_0-g_vz)^2-(m\pm g_sz)^2-p_\bot^2-\sqrt{g_s^2-g_v^2}\right]
\begin{pmatrix}
\tilde\varphi^\pm_1 \\
\tilde\varphi^\pm_2 \\
\end{pmatrix},\nonumber\\
&& \left[{d^2\over dz^2}+(p_0-g_vz)^2-(m\pm g_sz)^2-p_\bot^2+\sqrt{g_s^2-g_v^2}\right]
\begin{pmatrix}
\tilde\varphi^\pm_3 \\
\tilde\varphi^\pm_4 \\
\end{pmatrix}.
\end{eqnarray}

By introducing the dimensionless variable
\begin{equation}
    \zeta_\pm=(g_s^2-g_v^2)^{1/4}\left(z+\frac{g_v p_0\pm g_s m}{g_s^2-g_v^2}\right),
\end{equation}
and the dimensionless parameter
\begin{equation}
    \Delta_\pm=(g_s^2-g_v^2)^{-1/2}\left(\frac{(g_v p_0 \pm g_s m)^2}{g_s^2-g_v^2}+p_0^2-p_\bot^2-m^2\right),
\end{equation}
the two equations for the components are simplified as
\begin{eqnarray}
&&  \left(\frac{d^2}{d\zeta_\pm^2}-\zeta_\pm^2+\Delta_\pm-1\right)
    \begin{pmatrix}
        \tilde\varphi^\pm_1 \\
        \tilde\varphi^\pm_2 \\
    \end{pmatrix}
    =0, \nonumber\\
&&  \left(\frac{d^2}{d\zeta_\pm^2}-\zeta_\pm^2+\Delta_\pm+1\right)
    \begin{pmatrix}
        \tilde\varphi^\pm_3 \\
        \tilde\varphi^\pm_4 \\
    \end{pmatrix}
    =0.
\end{eqnarray}
While the two equations look similar to the complex equation (\ref{fxi}) in the case with only electric potential, the solutions here are real parabolic cylinder functions,
\begin{eqnarray}
&& \tilde\varphi^\pm_1,\ \tilde\varphi^\pm_2 \sim D_{\Delta_\pm/2-1}(\pm\sqrt 2\zeta_\pm),\nonumber\\
&& \tilde\varphi^\pm_3,\ \tilde\varphi^\pm_4 \sim D_{\Delta_\pm/2}(\pm\sqrt 2\zeta_\pm).
\end{eqnarray}
Through the transformation (\ref{trans}), a general solution $\varphi^\pm(z)$ of the second-order Dirac equation (\ref{varphi}) can be expressed in terms of the real parabolic cylinder functions,
\begin{eqnarray}
&& \varphi^\pm(z) = \begin{pmatrix}
        iP_\pm C_2^\pm D_{\Delta_\pm/2-1}(\pm\sqrt 2\zeta_\pm)-iP_\pm C_4^\pm D_{\Delta_\pm/2}(\pm\sqrt 2\zeta_\pm) \\
        -iP_\pm C_1^\pm D_{\Delta_\pm/2-1}(\pm\sqrt 2\zeta_\pm)+iP_\pm C_3^\pm D_{\Delta_\pm/2}(\pm\sqrt 2\zeta_\pm) \\
        C_2^\pm D_{\Delta_\pm/2-1}(\pm\sqrt 2\zeta_\pm)+C_4^\pm D_{\Delta_\pm/2}(\pm\sqrt 2\zeta_\pm) \\
        C_1^\pm D_{\Delta_\pm/2-1}(\pm\sqrt 2\zeta_\pm)+C_3^\pm D_{\Delta_\pm/2}(\pm\sqrt 2\zeta_\pm) \\
    \end{pmatrix}
\end{eqnarray}
with the eight constants $C^\pm_i\ (i=1,2,3,4)$ to be determined by the boundary condition. Substituting the known $\varphi^\pm$ into the equation (\ref{Dirac4}) for $\psi(t,{\mathbf x})$, taking the matrix calculations in the spin space for $\gamma^\mu$, and using the recursion relations for $D_\nu(z)$, which are similar to the treatment shown in the last section, a tedious but straightforward computing leads to a general solution of the Dirac equation (\ref{Dirac4}) which can still be expressed in terms of $\varphi^\pm$, but the coefficients $C_i^\pm$ should satisfy some constraint relations, which can be explicitly written as
\begin{eqnarray}
\label{Ci1}
&& C_1^+ +a_+ C_3^+ +b_+ C_4^+=0,\nonumber \\
&& C_2^+ -b_- C_3^+ +a_+ C_4^+=0,\nonumber \\
&& C_1^- +a_- C_3^- -b_+ C_4^-=0,\nonumber \\
&& C_2^- +b_- C_3^- +a_- C_4^-=0
\end{eqnarray}
with
\begin{equation}
a_\pm=\frac{g_s p_0 \pm g_v m}{\sqrt 2(g_s^2-g_v^2)^{3/4}},\ \ \ \ \ b_\pm=\frac{ip_\pm}{\sqrt 2(g_s^2-g_v^2)^{1/4}}.
\end{equation}

Using the asymptotic behavior for the real parabolic cylinder functions,
\begin{equation}
\lim_{\zeta_\pm \to \pm\infty}D_{\Delta_\pm/2}(\pm\sqrt 2\zeta_\pm)\ \sim \lim_{\zeta_\pm \to \pm\infty}D_{\Delta_\pm/2-1}(\pm\sqrt 2\zeta_\pm) \sim e^{-\zeta_\pm^2/2},
\end{equation}
it is clear that the above solution $\varphi^\pm(z)$ is a bound state with the boundary condition
\begin{equation}
\lim_{z\to\pm\infty}\varphi^\pm(z)=0.
\end{equation}
The coefficients $C_i^\pm$ are constrained by not only the equations (\ref{Ci1}) but also the continuous condition at the connection point of $\varphi^+$ and $\varphi^-$ at $z=0$,
 \begin{equation}
     \varphi^+(z=0^+)=\varphi^-(z=0^-),
 \end{equation}
which means
\begin{eqnarray}
\label{Ci2}
&& c_+ C_2^+ -d_+ C_4^+ -c_- C_2^- +d_- C_4^- = 0,\nonumber\\
&& c_+ C_1^+ -d_+ C_3^+ -c_- C_1^- +d_- C_3^- = 0,\nonumber\\
&& c_+ C_2^+ +d_+ C_4^+ +P_+^2c_- C_2^- +P_+^2d_- C_4^- = 0,\nonumber\\
&& c_+ C_1^+ +d_+ C_3^+ +P_+^2c_- C_1^- +P_+^2d_- C_3^- = 0
\end{eqnarray}
with
\begin{eqnarray}
&& c_\pm =iP_\pm D_{\Delta_\pm/2-1}(\pm\sqrt 2\zeta_\pm^0),\ \ \ \ \ d_\pm =iP_\pm D_{\Delta_\pm/2}(\pm\sqrt 2\zeta_\pm^0),\nonumber\\
&& \zeta_\pm^0 = {g_v p_0\pm g_s m\over (g_s^2-g_v^2)^{3/4}}.
\end{eqnarray}
To derive the last two equations of (\ref{Ci2}), we have used the condition $P_+ P_-=-1$.

\begin{figure}[h!]
	\centering
	\includegraphics[width=0.55\textwidth]{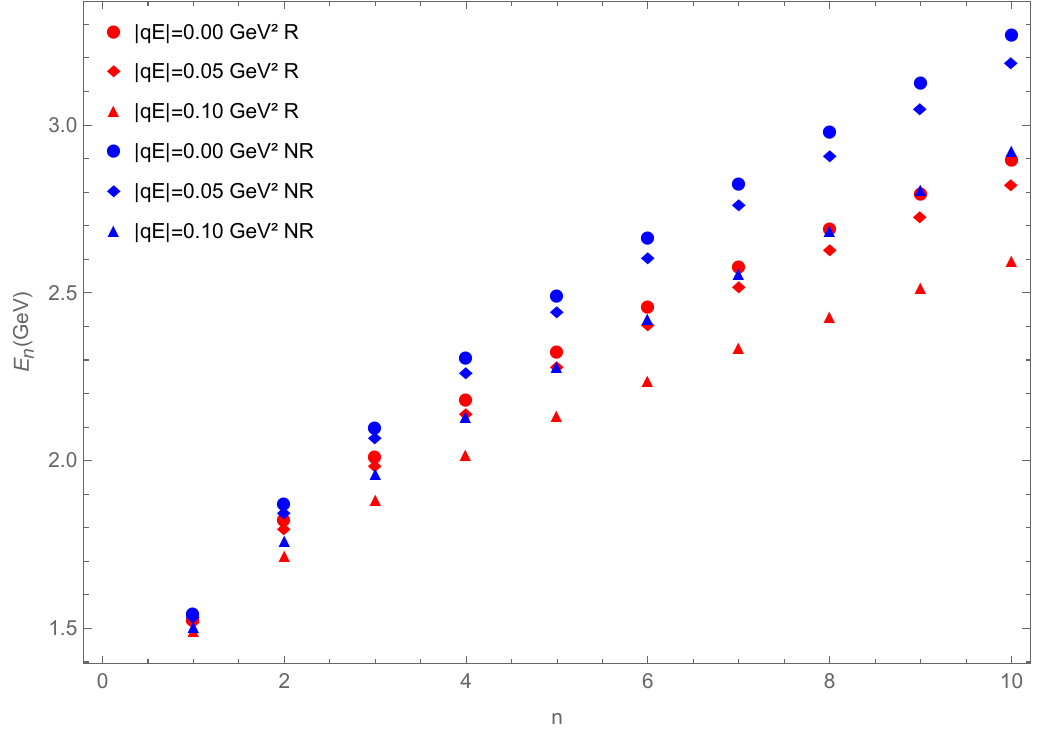}
	\caption{The lowest 10 energy levels for confined charm quarks in an external electric field with strength $|qE|=0, 0.05, 0.1$ GeV$^2$. $R$ and $NR$ indicate the solutions from the Dirac equation and   Schr\"odinger equation. }
	\label{fig1}
\end{figure}

\begin{figure}[h!]
	\centering
	\includegraphics[width=0.55\textwidth]{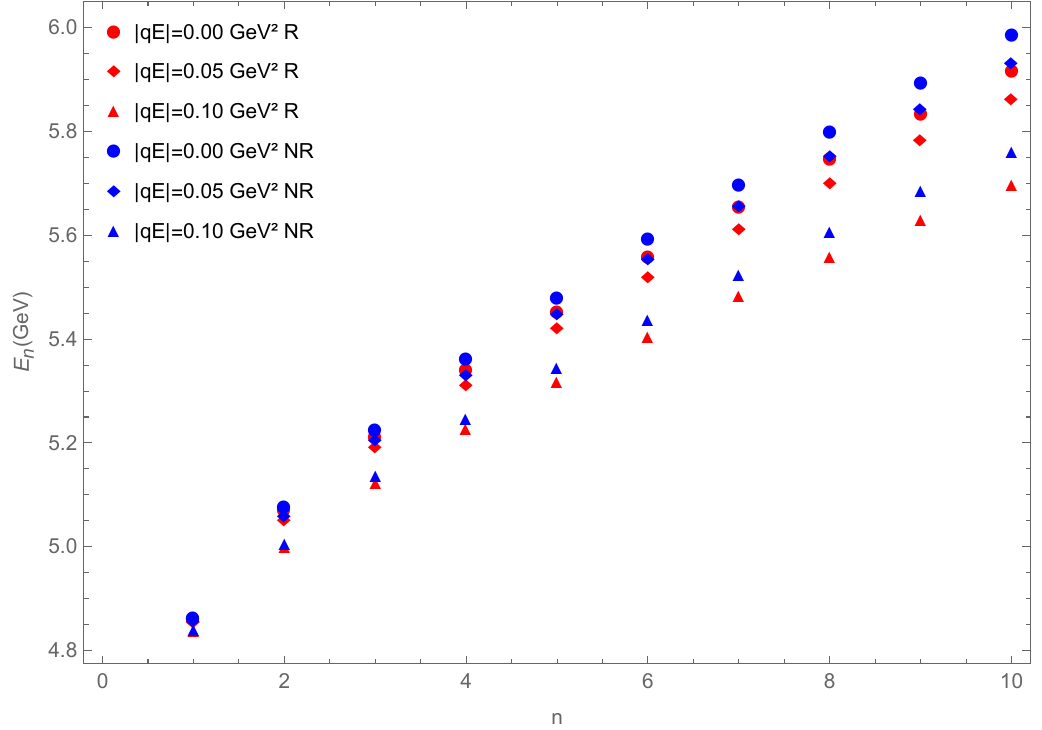}
	\caption{The lowest 10 energy levels for confined bottom quarks in an external electric field with strength $|qE|=0, 0.05, 0.1$ GeV$^2$. $R$ and $NR$ indicate the solutions from the Dirac equation and  Schr\"odinger equation. }
	\label{fig2}
\end{figure}

A non-trivial solution for the coefficients $C_i^\pm$ from their linear equations (\ref{Ci1}) and (\ref{Ci2}) requires a zero determinant of their coefficient matrix,
\begin{equation}
\left|
\begin{matrix}
       1&         0&       a_+&       b_+&          0&          0&          0&          0\\
       0&         1&      -b_-&       a_+&          0&          0&          0&          0\\
       0&         0&         0&         0&          1&          0&        a_-&       -b_+\\
       0&         0&         0&         0&          0&          1&        b_-&        a_-\\
       0&       c_+&         0&      -d_+&          0&       -c_-&          0&        d_-\\
    c_+&         0&       -d_+&         0&        -c_-&          0&       d_-&          0\\
       0&  c_+&         0&  d_+&          0&  P_+^2c_-&          0&  P_+^2d_-\\
c_+&         0&  d_+&         0&  P_+^2c_-&          0&  P_+^2d_-&          0\\
\end{matrix}\right | =0.
\end{equation}
This is the equation to determine the fermion energy spectrum $p_0=E_n$. As an example, we consider confined heavy quarks in an electric field. Since heavy quarks are so heavy, their motion can be approximately described by quantum mechanics, namely the Dirac equation or even the Schr\"odinger equation. We take charm quark mass $m_c=1.29$ GeV, bottom quark mass $m_b=4.7$ GeV, and scalar coupling constant $g_s=0.2$ GeV$^2$~\cite{Zhao:2020jqu}. Considering the fact that the strongest electric field created in relativistic heavy ion collisions is about $|eE|=0.1$ GeV$^2$, and the quark charge is less than the electron charge $|q|<|e|$, the condition $g_s > g_v$ is guaranteed in the calculation. For $p_x=p_y=0$, the lowest $10$ energy levels are shown in Fig.\ref{fig1} for charm quarks and Fig.\ref{fig2} for bottom quarks.
%%%%%%%%%%%%%%%%%%%%%%%%%%%%%%%%%%%%%%%%%%%%%%%%%%%%%
%%%%%%%%%%%%%%%%%%%%%%%%%%%%%%%%%%%%%%%%%%%%%%%%%%%%%

To compare the results with the non-relativistic limit, we now proceed to solve the corresponding Schr\"odinger equation,
\begin{equation}
    \left(m-\frac{1}{2m}\frac{d^2}{dz^2}+g_vz+g_s|z|\right)\phi(z)=p_0\phi(z).
\end{equation}
In order to effectively compare the two cases, we have shifted the non-relativistic energy by the particle mass $m$. It is well known that the solution of the equation is the Airy function $Ai(u)$,
\begin{equation}
\phi(z)=\left\{
\begin{aligned}
A_+ Ai\left((2mg_+)^{1/3}\left(z-\frac{p_0-m}{g_+}\right)\right),\ \ \ \ \ z>0\\
A_- Ai\left(-(2mg_-)^{1/3}\left(z+\frac{p_0-m}{g_-}\right)\right),\ \ \ \ \ z<0 \\
\end{aligned}\right.
\end{equation}
with $g_\pm=g_s\pm g_v$. The solution guarantees the boundary condition
\begin{equation}
\lim_{z\to\pm\infty}\phi(z)=0.
\end{equation}

Considering the continuous conditions for $\phi(z)$ and its first-order derivative $\phi'(z)=d\phi(z)/dz$ at $z=0$,
 \begin{eqnarray}
&&     \phi(z=0^+)=\phi(z=0^-),\nonumber\\
&&     \phi'(z=0^+)=\phi'(z=0^-),
\end{eqnarray}
namely,
\begin{eqnarray}
&& A_+ Ai\left(-(2mg_+)^{1/3}\frac{p_0-m}{g_+}\right) = A_- Ai\left(-(2mg_-)^{1/3}\frac{p_0-m}{g_-}\right),\nonumber\\
&& A_+ (2mg_+)^{1/3}Ai'\left(-(2mg_+)^{1/3}\frac{p_0-m}{g_+}\right) = -A_-(2mg_-)^{1/3}Ai'\left(-(2mg_-)^{1/3}\frac{p_0-m}{g_-}\right),
\end{eqnarray}
the non-relativistic energy $\epsilon$ is controlled by the determinant,
\begin{equation}
\left|\begin{matrix}
 Ai\left(-(2mg_+)^{1/3}\frac{p_0-m}{g_+}\right)&  -Ai\left(-(2mg_-)^{1/3}\frac{p_0-m}{g_-}\right)  \\
(2mg_+)^{1/3}Ai'\left(-(2mg_+)^{1/3}\frac{p_0-m}{g_+}\right)& (2mg_-)^{1/3}Ai'\left(-(2mg_-)^{1/3}\frac{p_0-m}{g_-}\right)  \\
\end{matrix} \right| =0.
\end{equation}

The comparison between the non-relativistic and relativistic solutions is shown in Figs.\ref{fig1} and \ref{fig2}. Firstly, both eigenvalues get reduced by the external electric field, indicating a partial cancellation of the confinement by the electric field. A stronger electric field means a stronger cancellation. Secondly, at any electric field strength, the energy derived from the Dirac equation is lower than that from the Schr\"odinger equation. This negative relativistic correction can be understood from the change in the Hamiltonian. Neglecting the quark spin, the relativistic Hamiltonian can be approximately written as
\begin{equation}
    H_R=\sqrt{{\mathbf p}^2+m^2}+V({\mathbf r}) \approx m+{{\mathbf p}^2\over 2m}-{{\mathbf p}^4\over 8m^3}+V({\mathbf r}) = H_{NR}-{{\mathbf p}^4\over 8m^3}<H_{NR}.
\end{equation}

Finally, we simply consider the case with $g_v > g_s$. From the second-order Dirac equation (\ref{varphi}) in the limit of $z\to \pm\infty$,
\begin{equation}
\left[{d^2\over dz^2}+(g_v^2-g_s^2)z^2\right]\varphi^\pm(z)=0,
\end{equation}
$\varphi^\pm(z)$ behaves like
\begin{equation}
\lim_{z\to\pm\infty}\varphi^\pm(z) \sim e^{+\frac{i}{2}\sqrt{g_v^2-g_s^2}z^2},e^{-\frac{i}{2}\sqrt{g_v^2-g_s^2}z^2}.
\end{equation}
This means that, when the electric coupling is stronger than the confinement coupling $g_v > g_s$, the wave function oscillates even at infinity, the confinement is fully canceled by the electric field, and the fermion is in a deconfinement state.

%%%%%%%%%%%%%%%%%%%%%%%%%%%%%%%%%%%%%%%%%%%%%%
\section{Bound state in a finite system}
\label{sec4}
%%%%%%%%%%%%%%%%%%%%%%%%%%%%%%%%%%%%%%%%%%%%%%
We consider now the possibility of having a bound state in a finite system. Suppose the fermion moving in an external electric field is restricted in a region with finite length $2L$ in the $z$-direction, \begin{equation}
\Omega=\left\{-\infty < x, y < +\infty, -L < z < +L\right\}.
\end{equation}
To have a bound state in $\Omega$, appropriate boundary conditions should be imposed on the surface of the region $\partial\Omega$ to ensure the energy quantization of the system. From the definition for the inner product of two solutions of the Dirac equation $\chi$ and $\eta$ in coordinate representation,
\begin{equation}
(\chi,\eta)=\int_\Omega d^3{\mathbf x}\chi^\dagger({\mathbf x})\eta({\mathbf x}),
\end{equation}
the requirement that the Hamiltonian (\ref{hamil}) of the system should be a hermitian operator, $(\chi,\hat H\eta) = (\hat H\chi,\eta)$, leads to the boundary condition
\begin{equation}
\chi^\dagger{\mathbf n}\cdot\gamma^0{\bm{\gamma}}\eta=0,\ \ \ \ \ \forall{\mathbf x}\in\partial\Omega,
\end{equation}
where ${\mathbf n}$ is the unit normal pointing outward to the boundary surface $\partial\Omega$. A trivial solution to this condition is the disappearing wave functions at the boundary, $\chi(z\to\pm\infty)=\eta(z\to\pm\infty)=0$, but from the conclusion of Section \ref{sec2}, this is forbidden in an external electric field. To look for a non-trivial solution, we impose the same boundary condition on $\chi$ and $\eta$,
\begin{equation}
\label{hermi}
\chi=K\chi,\ \ \ \ \ \eta=K\eta,\ \ \ \ \ \forall {\mathbf x}\in\partial\Omega,
\end{equation}
where $K$ is a $4\times4$ matrix which is determined by two conditions
\begin{equation}
\label{K1}
K^2=\mathbb{I}_4
\end{equation}
and
\begin{equation}
\label{K2}
K^\dagger\gamma^0{\mathbf n}\cdot{\bm \gamma}K=-\gamma^0{\mathbf n}\cdot{\bm \gamma}.
\end{equation}
The choice of the boundary condition (\ref{hermi}) leads to a vanishing probability current ${\mathbf j}$ in the direction normal to the boundary surface,
\begin{equation}
{\mathbf n}\cdot{\mathbf j}({\mathbf x}) = \psi^\dagger({\mathbf x}){\mathbf n}\cdot\gamma^0{\bm \gamma}\psi({\mathbf x}) = \psi^\dagger({\mathbf x})K^\dagger{\mathbf n}\cdot\gamma^0{\bm \gamma}K\psi({\mathbf x})
= -\psi^\dagger({\mathbf x}){\mathbf n}\cdot\gamma^0{\bm \gamma}\psi({\mathbf x}) = 0,\ \ \ \ \ \forall{\mathbf x}\in\partial\Omega.
\end{equation}
While the probability density $\rho({\mathbf x})=\psi^\dagger({\mathbf x})\psi({\mathbf x})$ itself does not vanish at the boundary, the probability current cannot flow outside the region. Under this boundary condition, the particle moving within the region $\Omega$ cannot penetrate the boundary surface, as it is confined to $\Omega$.

It is easy to check that
\begin{equation}
K=-i{\mathbf n}\cdot{\bm \gamma}
\end{equation}
satisfies the two conditions (\ref{K1}) and (\ref{K2}), it leads to the well-known MIT bag boundary condition~\cite{Chodos:1974je}
\begin{equation}
\label{MIT}
\chi= -i{\mathbf n}\cdot{\bm \gamma}\chi,\ \ \ \ \ \eta= -i{\mathbf n}\cdot{\bm \gamma}\eta,\ \ \ \ \ \forall{\mathbf x}\in\partial\Omega.
\end{equation}
For our choice of $\Omega$, the boundary condition (\ref{MIT}) leads to the constraint on the wave function $\phi(z)$,
\begin{equation}
\label{boun}
(1\pm i\gamma^3)\phi(z=\pm L)=0.
\end{equation}

From Section \ref{sec2}, a general wave function $\phi(z)$ is a linear superposition of the four independent wave functions $\phi_-^s(z,\pm)$. Applying the boundary condition (\ref{boun}) to it, and taking into account the fermion motion in the $z$-direction ($p_x=p_y=0$), a tedious but straightforward calculation leads to the energy spectrum governed by the equation,
\begin{eqnarray}
0&=& (1+i)m\sqrt{|qE|}\left[D_{i\lambda/2-1}(\xi_+(-L))D_{i\lambda/2}(-\xi_+(L))+D_{i\lambda/2-1}(-\xi_+(-L))D_{i\lambda/2}(\xi_+(L))\right]\nonumber\\
&& +m^2\left[D_{i\lambda/2-1}(\xi_+(-L))D_{i\lambda/2-1}(-\xi_+(L))-D_{i\lambda/2-1}(-\xi_+(-L))D_{i\lambda/2-1}(\xi_+(L))\right]\nonumber\\
&& +(1+i)m\sqrt{|qE|}\left[D_{i\lambda/2-1}(\xi_+(L))D_{i\lambda/2}(-\xi_+(-L))+D_{i\lambda/2-1}(-\xi_+(L))D_{i\lambda/2}(\xi_+(-L))\right]\nonumber\\
&& +(1+i)^2|qE|\left[D_{i\lambda/2}(\xi_+(-L))D_{i\lambda/2}(-\xi_+(L))-D_{i\lambda/2}(-\xi_+(-L))D_{i\lambda/2}(\xi_+(L))\right].
\end{eqnarray}
For highly excited states with large energy $p_0$, the equation can be asymptotically expressed as
\begin{eqnarray}
0 &=& {(1+i)m\sqrt{2\pi |qE|}\over \Gamma(1-im^2/(2|qE|))}\left(\xi(+L)\xi(-L)\right)^{-im^2/(2|qE|)}e^{-2i p_0 L}\nonumber \\
&&\times \left[\left(-\xi(L)\right)^{im^2/|qE|}e^{4i p_0 L}+\left(-\xi(-L)\right)^{im^2/|qE|}\right].
\end{eqnarray}
In the limit of $p_0\gg |qE|L$, the energy equation is further simplified to
\begin{equation}
\left({p_0\over \sqrt{|qE|}}\right)^{im^2/|qE|}\left(e^{4i p_0 L}+1\right)=0,
\end{equation}
which leads to an electric-field and particle-mass independent solution
\begin{equation}
p_0 = E_n = {(1+2n)\pi\over 4L},\ \ \ \ \ n\in\mathbb{Z}.
\end{equation}

%%%%%%%%%%%%%%%%%%%%%%%%%%%%%%%%%%%%%%%%%%%%%%
\section{Summary}
\label{sec5}
%%%%%%%%%%%%%%%%%%%%%%%%%%%%%%%%%%%%%%%%%%%%%%
In this work, we have presented a relativistic analysis of fermions in an external electric field by deriving the stationary solutions of the Dirac equation in infinite and finite systems. We strictly proved that bound states cannot exist in an infinite uniform electric field, based on the asymptotic expansion of the solutions. When a scalar confinement interaction is included, the bound state can survive in a weak electric field. As the field strength increases, however, the vector electric interaction eventually dominates, leading to the disappearance of the bound state. We analyzed also the  confinement enforced by the MIT bag–like boundary condition. By imposing a vanishing normal component of the probability current at the boundary, a stationary bound state can exist in a finite system, in contrast to the case of infinite volume. The energy spectrum is determined by the boundary condition.

The solutions of the Dirac equation provide the mode basis required to calculate QCD and QED Feynman diagrams in the presence of a background electric field. Considering that strong electromagnetic fields are generated in the early stage of relativistic heavy ion collisions, the Dirac spinors obtained here can be used to compute the dynamic processes like heavy quark pair production and Drell-Yan process in the collisions.

%\vspace{1cm}
\noindent {\bf Acknowledgement}: We thank Yantai University and SCNT for the financial support.

\appendix
\section{Properties of the Parabolic Cylinder Functions}
\label{app:cylinder}

In this appendix, we briefly review the definition and key properties of the parabolic cylinder functions $D_\nu(z)$ utilized in the text. The parabolic cylinder function $D_\nu(z)$ is a standard solution to the second-order linear ordinary differential equation, frequently referred to as Weber's equation~\cite{gradshteyn2014table}:
\begin{equation}
    \frac{d^2 y}{dz^2} + \left(\nu + \frac{1}{2} - \frac{z^2}{4}\right)y = 0.
\end{equation}
In our case, the equation governing the spatial part of the Dirac spinors, such as Eq.~(\ref{fxi}), can be solved analytically in terms of parabolic cylinder functions.

For a given parameter $\nu$, the two linearly independent solutions can be chosen as $D_\nu(z)$ and $D_\nu(-z)$, provided $\nu$ is not a non-negative integer. Their linear independence is verified by the Wronskian~\cite{bateman1953higher},
\begin{equation}
    D_\nu(z)\frac{d}{dz}D_\nu(-z)-D_\nu(-z)\frac{d}{dz}D_\nu(z)=\frac{\sqrt{2\pi}}{\Gamma{(-\nu)}}.
\end{equation}

In deriving the explicit matrix form of the spinor components, we have extensively applied the following recursion formulas to raise or lower the index $\nu$~\cite{gradshteyn2014table},
\begin{eqnarray}
&& D_{\nu+1}(z)-zD_\nu(z)+\nu D_{\nu-1}(z)=0,\nonumber\\
&& \frac{d}{dz}D_\nu(z)+\frac{1}{2}zD_\nu(z)-\nu D_{\nu-1}(z)=0,\nonumber\\
&& \frac{d}{dz}D_\nu(z)-\frac{1}{2}zD_\nu(z)+D_{\nu+1}(z)=0.
\end{eqnarray}

Finally, to determine the existence of bound states, it is necessary to examine the behavior of the wave functions in the limit of $z \to \pm \infty$. The asymptotic expansion of $D_\nu(z)$ for large arguments $|z|\gg 1$ and $|z|\gg |\nu|$ strictly depends on the region of $z$~\cite{gradshteyn2014table}:
\begin{eqnarray}
&&    D_\nu(z)\sim e^{-z^2/4}z^\nu\ \ \ \text{for}\ \ \ |arg(z)| < \frac{3}{4}\pi,\nonumber\\
&&    D_\nu(z)\sim e^{-z^2/4}z^\nu-\frac{\sqrt{2\pi}}{\Gamma(-\nu)}e^{i\pi\nu}e^{z^2/4}z^{-\nu-1}\ \ \ \text{for}\ \ \ \frac{1}{4}\pi<arg(z)<\frac{5}{4}\pi,\nonumber\\
&&    D_\nu(z)\sim e^{-z^2/4}z^\nu-\frac{\sqrt{2\pi}}{\Gamma(-\nu)}e^{-i\pi\nu}e^{z^2/4}z^{-\nu-1}\ \ \ \text{for}\ \ \ -\frac{1}{4}\pi>arg(z)>-\frac{5}{4}\pi.
\end{eqnarray}

\bibliographystyle{apsrev4-2}
\bibliography{main}

\end{document}